\documentclass[aps,prb,reprint,twocolumn,amsmath,longbibliography]{revtex4-2}
\usepackage{amsmath,amssymb,bm,mathrsfs,graphicx, braket, times,amsthm,enumerate, scrextend}
\usepackage[colorlinks=true,citecolor=blue,linkcolor=blue]{hyperref}
\usepackage{multirow}
\usepackage{array}
\usepackage{bigstrut}
\usepackage[all,cmtip]{xy}
\usepackage[normalem]{ulem}
\usepackage[usenames,dvipsnames]{color}
\usepackage{makecell}
\usepackage{xcolor}

\usepackage{threeparttable}
\usepackage{float}
\usepackage{booktabs}
\usepackage{subcaption}
\usepackage{stackengine}

\usepackage{ragged2e}
\usepackage{diagbox}

\begin{document}
\title{Large Berry curvature effects induced by extended nodal structures: Rational design strategy and high-throughput materials predictions}
\author{Wencheng Wang$^{1,2}$}
\author{Minxue Yang$^1$}
\author{Wei Chen$^1$}
\author{Xiangang Wan$^{1,3,4}$}
\author{Feng Tang$^1$}\email{fengtang@nju.edu.cn}

\affiliation{$^1$National Laboratory of Solid State Microstructures and School of Physics, Nanjing University, Nanjing 210093, China and Collaborative Innovation Center of Advanced Microstructures, Nanjing University, Nanjing 210093, China}
\affiliation{$^2$International Quantum Academy, Shenzhen 518048, China}
\affiliation{$^3$Hefei National Laboratory, Hefei 230088, China}
\affiliation{$^4$Jiangsu Physical Science Research Center}
\date{\today}

\begin{abstract}
	Berry curvature can drastically modify the electron dynamics, thereby offering an effective pathway for electron manipulation and novel device applications. Compared to zero-dimensional nodal points in Weyl/Dirac semimetals, higher-dimensional extended nodal structures, such as nodal lines and nodal surfaces, are more likely to intersect the Fermi surface, leading to large Berry curvature effects without fine-tuning the chemical potential. In this work, we propose a strategy that utilizes straight nodal lines (SNLs) and flat nodal surfaces (FNSs) to design large Berry curvature effects, and we exhaustively tabulate SNLs and FNSs within the 1651 magnetic space groups (MSGs). We demonstrate that SNLs and FNSs can generate large Berry curvature widely distributed in the Brillouin zone. As an application, we identify 158 MSGs that host FNSs, SNLs, or both and allow for nonvanishing anomalous Hall conductivity (AHC). Based on these 158 MSGs, we screen materials from the MAGNDATA magnetic material database for high-throughput calculations, identifying 60 materials with AHC values exceeding $500\,\Omega^{-1}{\rm cm}^{-1}$. We select the candidate materials $\rm SrRuO_3$ and $\rm Ca_2NiOsO_6$ to demonstrate the contributions of FNSs and SNLs to one and two nonvanishing AHC components, respectively. We also investigate the tuning of AHC through symmetry breaking, outlining all possible symmetry-breaking pathways, and select the candidate material HoNi to demonstrate this approach by applying an external magnetic field. Additionally, we identify Berry curvature quadrupoles in the candidate materials, indicating that our strategy can be generalized to Berry curvature multipole effects. Our work will guide both the theoretical and experimental design of materials with large Berry curvature effects, with significant implications for a wide range of device applications.   \par 
	
\end{abstract}

\maketitle

\section{Introduction}
The Berry curvature plays a crucial role in various physical effects~\cite{Berryphase-Niu-rmp}, including electric polarization~\cite{polar-1993-prb,polar-1994-rmp,polar-1994-prb}, orbital magnetization~\cite{orbmag-Thonhauser-prl,orbmag-Xiao-prl,orbmag-Shi-prl}, quantum charge pumping~\cite{pump-Niu-prl}, and all types of Hall effects~\cite{QHE-Klitzing-prl,TKNN-1982,qahe-haldane-prl,qshe-kane-prl,valleygraphene-xiao-prl,ahe-MacDonald-prl,ahe-Nagaosa-JPS,ahe-Haldane-prl,nlhe-fu-prl} along with their thermoelectric counterparts~\cite{ANE-Xiao-PRL,ANE-Onoda-PRB,NLANE-Xiao-PRB,ANE-Mn3Sn-NP,ahe-Co2MnGa-np,NLANE-TMDC,NLANE-WTe2}. Specifically, the quantum anomalous Hall effect~\cite{qahe-haldane-prl,qahe-yu-sci,qahe-xue-sci,QAHE-MnBi2Te4-sci,QAHE-MnBi2Te4-prl} is characterized by a nonzero Chern number\textemdash the quantized integral of the Berry curvature\textemdash whereas the anomalous Hall effect (AHE)~\cite{ahe-MacDonald-prl,ahe-Nagaosa-JPS,ahe-Haldane-prl,ahe-nagaosa-rmp}, nonlinear AHE (NLAHE)~\cite{nlhe-fu-prl,nlhe-du-nc}, and their thermoelectric counterparts arise from the Berry curvature in partially filled bands. These effects can be significantly enhanced near topological nodal structures with large Berry curvature distributions, typically found in Weyl semimetals~\cite{Weyl-Wan-PRB,TaAs-whm-PRX,TaAs-exp-sci, ahe-liu-np,ahe-wenghm-nc,nlhe-fu-prl,nlhe-du-prl,nlhe-ma-nat,nlhe-WTe2-nm}, which have attracted tremendous attention over the past decade due to their fundamental significance and promising device applications. However, the material realization of such large Berry curvature effects remains challenging and is an ongoing pursuit. In particular, achieving large Berry curvature effects requires the zero-dimensional Weyl points to be close to the Fermi level~\cite{ahe-burkov-prl}, which has proven challenging in real materials~\cite{bc-fns-nature}. 
 
For the specific status of material realization, the most representative materials that exhibit large intrinsic AHE to date include magnetic Weyl semimetals $\rm Co_3Sn_2S_2$~\cite{ahe-liu-np,ahe-wenghm-nc} and $\rm Fe_3Sn_2$~\cite{ahe-Fe3Sn2-Nature}, Heusler compounds $\rm Co_2Mn\mathit{X} (\mathit{X}=Al,Ga)$~\cite{ahe-yan-nc,ahe-Co2MnGa-np}, chiral antiferromagnetic family $\rm Mn_3\mathit{X} (\mathit{X}=Sn, Ge)$~\cite{ahe-Mn3Sn-Nature,ahe-Mn3Ge-SA}, helimagnets $\rm Mn\mathit{X}(\mathit{X}=Si,Ge)$~\cite{ahe-MnSi-prb,ahe-MnGe-prl}, and van der Waals ferromagnet $\rm Fe_3GeTe_2$~\cite{ahe-Fe3GeTe2-nm}. Most of these materials require low temperatures to exhibit large AHC, so those with large AHC at temperatures above room temperature need further development for practical applications.   \par 

In addition to zero-dimensional nodal points, electronic bands can also host higher-dimensional extended nodal structures, such as nodal lines/loops~\cite{NLSM-Burkov-prb,NLSM-Fang-prb,SNLFNS-Liang,ring-Mao-PRL,hourglass-wu-prb,bcsg-wu-prb,bcmsg-tang-prb,bcsg-yao-scibull,bcmsg3-yao-prb,bcmsg4-yao-prb}, nodal links~\cite{Hopflink-Chen-PRB,Hopflink-Wang-PRB,nodalchain-Hasan,hopflink-exp-Hasan,Hopflink-Ezawa,ahe-yan-nc,bcsg-wu-prb}, nodal chains~\cite{chainnet-Nature,nodalchain-Hasan,chainnet-NP,bcsg-wu-prb}, nodal nets~\cite{nodalnet-Wang-PRL}, and nodal surfaces/planes~\cite{SNLFNS-Liang,NSSM-Zhong,NSSM-curved,NSSM-Yang-PRB,nodalsurface-wilde-nat,bcsg-wu-prb,bcmsg-tang-prb,bcsg-yao-scibull,bcmsg3-yao-prb,bcmsg4-yao-prb}. Owing to their extended nature, these nodal structures are more likely to intersect the Fermi surface, thereby enabling large Berry curvature effects without the need for fine-tuning, in contrast to the case of nodal points. Among them two particular types of extended nodal structures are of great interest: one-dimensional straight nodal lines (SNLs) and two-dimensional flat nodal surfaces (FNSs). The main advantage of these two types of nodal structures is that they not only span the entire Brillouin zone (BZ)~\cite{bc-fns-nature,bcquadrupole-chan-prl}, but also generally extend over a continues energy window. This suggests that SNLs and FNSs are much more likely to intersect the Fermi surface compared with the nodal points, both from the perspective of momentum space and energy scale, thereby leading to large Berry curvature effects. Given that existing studies on large AHC are mostly case-by-case, high-throughput investigations of large Berry curvature effects induced by SNLs and FNSs are highly desirable, as they would greatly benefit material selection and device applications. In addition, the AHC tuning is also of great interest and highly valuable for applications, with approaches such as chemical composition tuning~\cite{ahctune-chemi-prl,htc-chemitune-ActaM}, pressure tuning~\cite{ahctune-press-prm}, electric tuning~\cite{ahctune-elec-prl}, and magnetic tuning~\cite{ahe-yan-nc}. 

In this work, we propose a design strategy that utilizes the SNLs and FNSs to induce large Berry curvature effects, and we exhaustively tabulate all the possible SNLs and FNSs in the 1651 magnetic space groups (MSGs) using double-valued (single-valued) representations corresponding to spinful (spinless) system. Then, based on the tabulation of the SNLs and FNSs, we construct effective Hamiltonians for the selected SNLs and FNSs to investigate their Berry curvature distributions in the BZ. We demonstrate the advantages of using SNLs and FNSs to induce large Berry curvature effects. As an application, we investigate the symmetry-adapted AHC tensors for all 1651 MSGs, and highlight 158 MSGs that host FNSs, SNLs, or both and allow for nonvanishing AHC. Next, guided by these 158 MSGs, we screen materials from the MAGNDATA magnetic material database, and perform high-throughput first-principles calculations to identify materials exhibiting large AHC. Furthermore, we investigate the AHC tuning through symmetry breaking, and outline all possible symmetry-breaking pathways.    \par 

This paper is organized as follows. In Sec. \ref{design}, we present the theoretical design strategy and derived results. In Sec. \ref{FNSSNL}, we tabulate all possible SNLs and FNSs in 1651 MSGs and investigate the symmetry-adapted AHC tensor of these MSGs. In Sec. \ref{toymodel}, based on the  tabulation, we construct two effective Hamiltonians for representative SNLs and FNSs, and investigate their Berry curvature distributions in the BZ. We illurstrate that both SNLs and FNSs can exhibit broad energy window series with large Berry curvature effects. In Sec. \ref{HTC}, we perform theory-driven high-throughput calculations to search for magnetic materials with large AHC. In Sec. \ref{candidate1} and \ref{candidate2}, we select two candidate materials, $\rm SrRuO_3$ and $\rm Ca_2NiOsO_6$, as representatives to showcase the large AHC induced by the SNLs and FNSs. In Sec. \ref{tune}, we explore the tuning of AHC through symmetry breaking, and use candidate material $\rm HoNi$ to demonstrate the AHC tuning via an external magnetic field. The calculation method is provided in Sec. \ref{method}. Finally, Sec. \ref{conclusion} contains conclusion and discussion.   \par

\begin{figure}[ht]
	\centering
	\includegraphics[width=0.48\textwidth]{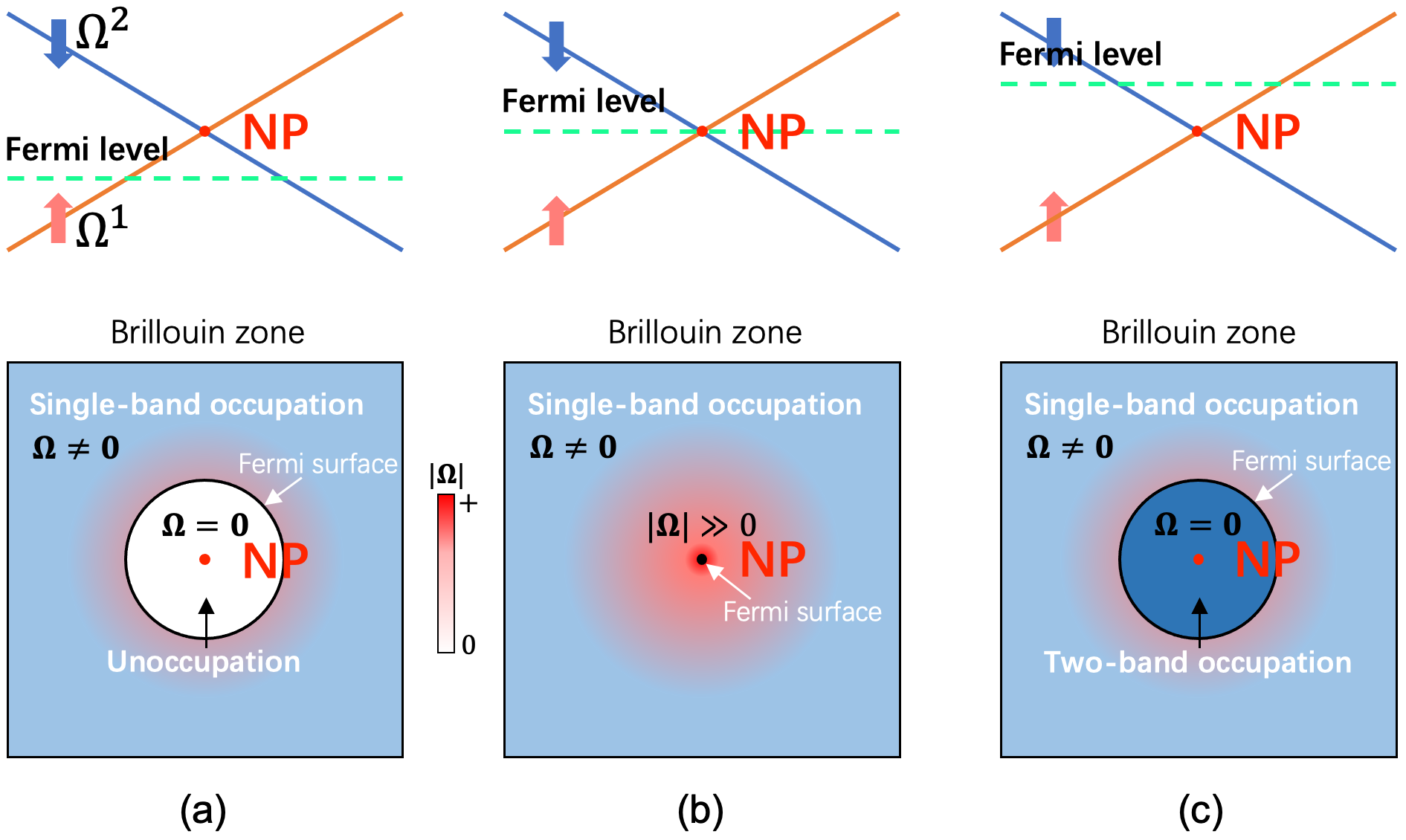}
	\captionsetup{justification=raggedright}
	\caption{Berry curvature distributions induced by a nodal point (NP) at different Fermi level positions. The occupied bands and their corresponding Berry curvature distributions are shown in the upper and lower panels, respectively. The opposite signs of the Berry curvature of the two bands are represented by red and blue arrows. (a) When the Fermi level is below the NP, the single-band occupied region with nonzero Berry curvature ${\bm\Omega} \neq {\bm 0}$ is shown in light blue, while the NP lies in the unoccupied region (white), where ${\bm\Omega} = {\bm 0}$. (b) When the Fermi level is at the NP, the entire BZ is single-band occupied, and large Berry curvature appears, concentrated at the NP. (c) When the Fermi level is above the NP, the NP lies in the two-band occupied region (dark blue), where ${\bm\Omega} = {\bm 0}$ due to the cancellation of Berry curvatures between the two bands. }
	\label{Fermi_level}
\end{figure}

\section{Theoretical design of large Berry curvature effects by extended nodal structures} \label{design}
We first conduct a theoretical analysis of the conditions for the large Berry curvatures effects such as AHE. For any given system, the intrinsic AHC $\sigma$ is proportional to the sum of the Berry curvature $\Omega_{ab}^n(\bm k)$, where $a,b\in \{x,y,z\}$ and $n$ denotes the band index, over all occupied states~\cite{ahe-nagaosa-rmp,nlhe-du-nrp}. This can be concisely expressed as: $\sigma_{ab} \propto \sum_{\bm k}\sum_nf_{n,\bm k} \Omega_{ab}^n(\bm k)$, where $f_{n,\bm k}$ is the Fermi-Dirac distribution function. The first condition is obviously that the energy bands should exhibit large Berry curvature distributions, which generally arise near the topological band touchings (BTs)~\cite{Berryphase-Niu-rmp}. Second, the local conservation law for the Berry curvature indicates that the strongest Berry curvatures occur when the cancellation between different bands is minimized. Taking a two-band system with a nodal point as an example, only k-points with single-band occupation contribute a net Berry curvature, as illustrated by the light blue regions in Fig. \ref{Fermi_level}. Therefore, the intrinsic AHE is strongest when the Fermi level coincides with the nodal point, corresponding to the weakest interband compensation of the Berry curvature. Therefore, the AHE reaches its maximum strength when the Fermi level approaches the nodal points. Third, the symmetry conditions must be satisfied for different physical scenarios~\cite{multipole-Watanabe,multipole-Yatsushiro,symcons-NSR}. For instance, time-reversal symmetry (TRS) and parity-time symmetry must be broken to obtain a finite AHE. Moreover, the properties of the system are further constrained by the crystal symmetries of the crystallographic groups, such as the space group (SG)~\cite{book-bradley-oxford,bilbao} and magnetic space group (MSG)~\cite{kpmodelmsg-tang-prb,M-TQC}. 

In summary, achieving a large Berry curvature effect requires satisfying three necessary conditions: (1) The presence of topological BTs with large Berry curvature distributions; (2) The Fermi level intersecting or approaching the BTs; (3) The fulfillment of additional symmetry constraints. It is worth noting that the above analysis is general, regardless of whether the system is non-magnetic or magnetic, the specific magnetic configuration, as well as the specific origin of the Berry curvature. These three necessary conditions are not easily satisfied simultaneously in most materials, thus posing challenges for material realization. However, conversely, they can also serve as guiding principles for designing materials with large intrinsic Berry curvature effects.  \par

\begin{table*}[ht]
	\renewcommand\arraystretch{1.25}
	\captionsetup{justification=raggedright}
	\caption{The 158 MSGs that host FNSs, SNLs, or both and allow for a nonvanishing AHC tensor. The three columns contain: the symmetry-adapted AHC tensors of the MSGs, the nodal structures hosted by these MSGs, and the corresponding MSGs. Notably, all 158 MSGs belong to type-III MSGs, and the table applies to both double- and single-valued representations. }
	\label{AHC_MSG}
	\begin{tabular}{p{0.15\textwidth} p{0.15\textwidth} p{0.65\textwidth}}
		
		\hline\hline
		
		\textbf{AHC tensor } & \textbf{Nodal structures} & \textbf{MSGs} (for both double- and single-valued representations)  \\
		\hline
		
		$(\sigma_{yz}, 0, 0)$ & Only SNLs &  30.114 39.197 41.213 66.496 67.506 68.516 72.544 74.559 \\ 
		& Only FNSs & 18.19 20.34 26.69 29.102 31.126 36.175 51.295 63.463 \\
		& SNLs and FNSs & 33.147 52.311 53.327 54.343 57.383 58.398 59.410 60.423 62.447 64.475 \\
		
		\hline
		$(0 ,\sigma_{zx},0)$ & Only SNLs &  27.80 28.89 30.113 32.137 34.158 37.182 39.199 40.207 41.215 43.226 \
		45.237 46.243 49.270 50.282 70.530 \\ 
		& Only FNSs &  17.10 26.68 36.174 51.296 63.464 \\
		& SNLs and FNSs & 29.101 31.125 33.146 52.312 53.328 54.344 55.358 56.370 57.384 60.424 61.436 62.448 64.476 \\
		
		\hline
		$(0 ,0 ,\sigma_{xy})$ & Only SNLs & 26.70 27.81 28.91 29.103 30.115 31.127 32.138 33.148 34.159 36.176 \
		37.183 40.206 41.214 43.227 45.238 46.245 48.260 49.269 50.281 51.294 \
		52.310 53.326 54.342 63.462 64.474 66.495 68.515 72.543 73.551 100.175 \
		101.183 102.191 103.199 104.207 105.215 106.223 108.237 109.243 \
		110.249 112.263 116.295 117.303 118.311 120.325 122.337 124.357 \
		125.369 126.381 131.441 132.453 133.465 134.477 140.547 141.557 \
		142.567 158.59 159.63 161.71 163.83 165.95 167.107 184.195 185.201 \
		186.207 188.219 190.231 192.250 193.260 194.270 \\
		& Only FNSs & 18.18 19.27 55.357 57.382 59.409 90.98 92.114 94.130 96.146 113.271 \
		127.393 129.417 \\
		&  SNLs and FNSs & 56.369 58.397 60.422 62.446 114.279 128.405 130.429 135.489 136.501 137.513 138.525  \\
		
		\hline
		$(\sigma_{yz}, \sigma_{zx}, 0)$ & Only SNLs & 7.26 9.39 13.69 15.89 \\
		& Only FNSs & 4.9 11.54 \\
		& SNLs and FNSs& 14.79 \\
		
		\hline\hline		
	\end{tabular}
\end{table*}

\subsection{Symmetry-enforced extended nodal structures and symmetry-adapted AHC tensors in the 1651 MSGs} \label{FNSSNL}

We aim to utilize the extended nodal structures, namely the SNL and FNS, as generators of large Berry curvature, thereby fulfilling the first necessary condition discussed above. We consider all possible SNLs and FNSs in the 1651 MSGs, which include 230 type-I MSGs (without TRS operation), 230 type-II MSGs (with TRS operation), 674 type-III MSGs (half SG operations multiplied by TRS operation), and 517 type-IV MSGs (all SG operations multiplied by TRS operation and an additional fractional translation). Then, based on the irreducible representations (irreps) or irreducible corepresentations (co-irreps) of the little group of high-symmetry lines (HSLs) and high-symmetry planes (HSPLs), along with their compatibility relations (CRs), we identify all possible SNLs and FNSs in these 1651 MSGs, which coincide with HSLs and HSPLs, respectively. Thus, these SNLs and FNSs are symmetry-enforced. The single-valued and double-valued irreps (co-irreps) apply to systems with negligible and significant SOC, respectively, as well as to spinless and spinful systems. Note that we follow the MSG convention adopted in Ref.~\cite{book-bradley-oxford} throughout this paper. 

The method for identifying SNLs and FNSs can be briefly summarized as follows. For a given HSL, if a degenerate irrep (co-irrep) of its little group exists, and the degenerate irrep (co-irrep) can decompose into more than one irrep (co-irrep) of all neighboring HSPLs and general points (GPs), the HSL can be an SNL. Similarly, for a given HSPL, if a degenerate irrep (co-irrep) of its little group exists, and it can decompose into more than one irrep (co-irrep) of all neighboring GPs, the HSPL can be an FNS. Furthermore, if this condition holds for all irreps (co-irreps) of the little group, the SNL or FNS necessarily exists. All results are provided in Supplemental Material (SM) I~\cite{SM1}. For each MSG, we tabulate all possible SNLs and FNSs with their concrete positions in the BZ. If an SNL or FNS necessarily exists, its position $\bm k$ is highlighted in red. The little group of each SNL or FNS is also provided, which can be used to analyze symmetry-adapted physical quantities at the SNL or FNS, such as Berry curvature and quantum metric. Additionally, several SNLs or FNSs can be related by symmetry operations of the MSG, which are also explicitly provided. The statistics of MSGs with different types of nodal structures are presented in SM I~\cite{SM1}. Through this exhaustive tabulation, several general conclusions can be deduced. For example, all FNSs are protected by $\Theta C_{2i}$ symmetry ($\Theta$ denotes TRS operation, and $C_{2i}$ denotes the two-fold rotation operation about the $i$-axis, where $i=x, y, z$). Therefore, type-I MSG can not host FNSs as they lack TRS operation $\Theta$. Notably, all FNSs in the 1651 MSGs necessarily exist, regardless of whether SOC is considered or neglected. Specifically, 254 MSGs host FNSs when SOC is considered, and 486 MSGs host FNSs when SOC is neglected. Regarding the SNLs in the 1651 MSGs, although not all of them necessarily exist, approximately 94\% of the SNLs necessarily exist when SOC is considered. A total of 810 MSGs host necessarily existing SNLs with SOC, of which roughly 97\% of the SNLs necessarily exist. Furthermore, 131 MSGs necessarily host coexisting SNLs and FNSs when SOC is considered. \par

Considering the third necessary condition, the symmetries in MSGs not only protect the nodal structures but also impose constraints on the physical properties of the system, including the AHC tensor. The AHC tensor $\sigma_{ij}$ can be expressed as in Eq. (\ref{AHC_def}), where $E_{j}$ denotes the driving electric field, $j_i$ denotes the response current, and the indices $i,j,a,b\in \{x,y,z\}$, indicating that the AHC tensor $\sigma_{ij}$ characterizes the linear response of the current to the driving electric field. In particular, the AHC tensor is a second-rank antisymmetric tensor ($\sigma_{ij}=-\sigma_{ji}$) with only three independent components, which can also be expressed as a pseudovector ${\bm \sigma}=(\sigma_{yz},\sigma_{zx},\sigma_{xy})$. When a symmetry operation is applied to the system, it is equivalent to applying to both the driving electric field $E_j$ and the response current $j_i$. In this case, both $E_j$ and  $j_i$ are transformed, but the tensor $\sigma_{ij}$ remains invariant, and Eq. (\ref{AHC_def}) still holds. Thus, the symmetry operation can impose constraints on the AHC tensor $\sigma_{ij}$, as expressed in Eq. (\ref{symm_const}), where $R$ denotes the matrix representation of the point operation part of the symmetry operation in the MSG, and $\eta(T)$ represents the parity of the physical quantity that contributes to AHC under TRS operation. For instance, the Berry curvature $\bm \Omega(\bm k)$, which contributes to AHC, has $\eta(T)=-1$. Therefore, for a given MSG, all nonvanishing components of the tensor $\sigma_{ij}$ can be determined. The symmetry-adapted AHC tensor for each MSG is then calculated using Eq. (\ref{symm_const}) with all symmetry operation matrices (or generating matrices), and the results are provided in SM I~\cite{SM1}. Additionally, the point operation part of the symmetry operations in an MSG form a magnetic point group (MPG), and 31 MPGs allow for a nonvanishing AHC tensor, as detailed in SM I~\cite{SM1}. 

Moreover, by combining the MSGs hosting FNSs, SNLs, or both with those allowing for nonvanishing AHC, we identify 158 MSGs that host FNSs, SNLs, or both and allow for nonvanishing AHC. These MSGs are categorized based on their AHC tensors and nodal structures, as shown in Table \ref{AHC_MSG}. Notably, Table \ref{AHC_MSG} applies to both single-valued and double-valued representations, thus covering both spinless and spinful systems, such as phononic and electronic systems. Among these, 151 MSGs allow for only one nonvanishing AHC component, while 7 MSGs allow for two nonvanishing AHC components. Furthermore, classified by nodal structure types, the numbers of MSGs that allow for nonvanishing AHC and host only SNLs, only FNSs, or both SNLs and FNSs are 96, 27, and 35, respectively.  \par 

\begin{equation} \label{AHC_def}
	j_i=\sigma_{ij}E_j
\end{equation}

\begin{equation} \label{symm_const}
	\sigma_{ij}=\eta(T)\sum_{a,b}R_{ia}R_{jb}\sigma_{ab}
\end{equation}

\begin{figure*}[ht]
	\centering
	\includegraphics[width=1\textwidth]{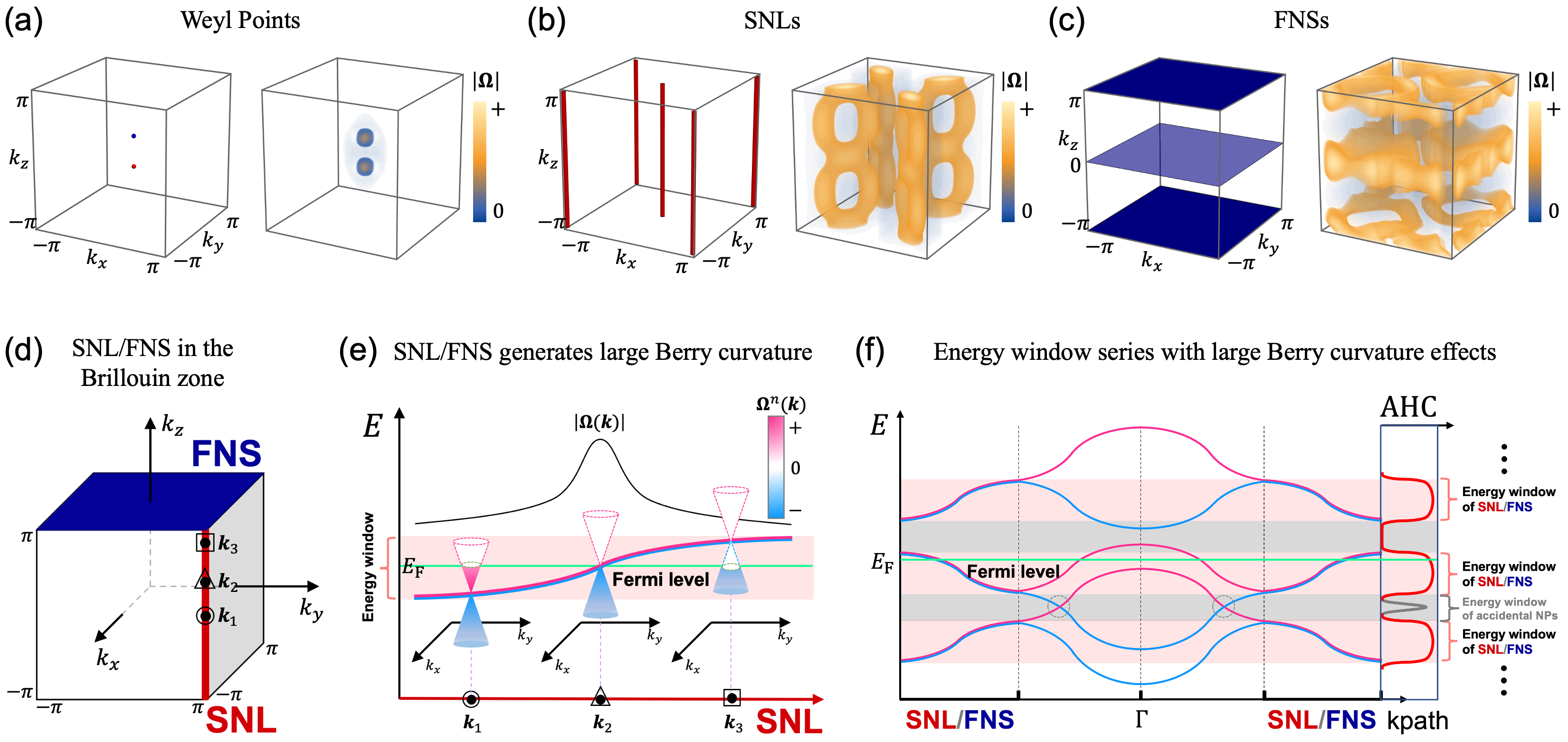}
	\captionsetup{justification=raggedright}
	\caption{Design of large Berry curvature effects by SNLs or FNSs. (a-c) Nodal structures and Berry curvature distributions of Weyl points, SNLs, and FNSs in the BZ. The effective Hamiltonians for the Weyl points, SNLs, and FNSs are given by Eq. (\ref{toy_weyl}), (\ref{toy_snl}), and (\ref{toy_fns}), respectively. (d) An SNL and an FNS coinciding with a HSL and a HSPL in the BZ, represented by a red line and a blue plane, respectively. Three representative k-points on the SNL are labeled as $\bm k_1$, $\bm k_2$, and $\bm k_3$. (e) The degenerate energy bands along the SNL cross the Fermi level, shown as overlapping red and blue curves. The three-dimensional energy bands are plotted in the $k_x$-$k_y$ plane at the three k-points, forming three Dirac cones. The Berry curvature at each k-point is determined by the portion of the cone below the Fermi level. (f) Energy window series with large Berry curvature effects induced by an necessarily existing SNL or FNS. All bands in such an SNL or FNS must be degenerate, leading to energy window series with large Berry curvature effects such as AHE. Two accidental NPs are indicated by grey dashed circles, which can contribute to the AHC in a relatively narrow energy window.  }
	\label{strategy}
\end{figure*}

\subsection{SNLs and FNSs generate widely distributed large Berry curvature and induce large AHE over broad energy windows} \label{toymodel}

Based on our comprehensive tabulation of SNLs and FNSs in 1651 MSGs, we select specific SNLs and FNSs as examples to demonstrate their Berry curvature distribution. In type-III MSG 81.35, the HSLs $(0,0,w)$ and $(\pi,\pi,w)$ necessarily host SNLs (where $\omega$ is real parameter), which are protected by the little group $\{E,C_{2z}, \Theta S_{4z}\}$. In type-III MSG 20.34, the HSPL $(u,v,\pi)$ necessarily hosts an FNS (where $u$ and $v$ are real parameters), which is protected by the little group $\{E,\Theta C_{2z}\}$. We then construct two effective Hamiltonians for these SNLs and FNSs, ensuring they satisfy the symmetries of their little groups. The effective Hamiltonians of these SNLs and FNSs are given in Eq. (\ref{toy_snl}) and (\ref{toy_fns}), respectively. The nodal structures of the SNLs and FNSs are depicted as red lines and blue planes in the left panels of Fig. \ref{strategy}(b) and (c), respectively. The Berry curvature distributions induced by these SNLs and FNSs in the BZ are calculated based on their effective Hamiltonians with half-filled bands ($E_{\rm F}=0\,{\rm eV}$), as shown in the right panels of Fig. \ref{strategy}(b) and (c). For comparison, we also derive the renowned effective Hamiltonian of the Weyl semimetal with two Weyl nodes at $(0,0,\pm k_0)$, where $k_0=\frac{\pi}{4}$, as expressed in Eq. (\ref{toy_weyl}). The Weyl points and Berry curvature distribution for half-filled bands are shown in the left and right panels of Fig. \ref{strategy}(a), respectively. It is evident that the large Berry curvature induced by Weyl points is highly concentrated at these Weyl points, while that induced by the SNLs and FNSs is more widely distributed throughout the BZ. Therefore, in momentum space, due to the extensibility of the SNLs and FNSs, they are more likely intersect the Fermi surface, resulting in both the Fermi surface and Fermi sea exhibiting large Berry curvature.   \par

~\\
\noindent
$H_{WP}={\sin}k_x\sigma_x+{\sin}k_y\sigma_y$
\begin{equation}\label{toy_weyl}
	+(2+{\cos}k_0-{\cos}k_x-{\cos}k_y-{\cos}k_z)\sigma_z
\end{equation}
$H_{SNL}={\sin}k_x{\sin}k_y\sigma_x+{\sin}^3k_x{\sin}^3k_y{\sin}k_z\sigma_y$
\begin{equation}\label{toy_snl}
	+({\cos}k_x-{\cos}k_y)\sigma_z
\end{equation}
$H_{FNS}=({\sin}k_x+{\sin}k_y){\sin}k_z\sigma_x+({\sin}k_x+{\sin}k_y){\sin}^2k_z\sigma_y$
\begin{equation}\label{toy_fns}
	+{\sin}k_z\sigma_z
\end{equation}

Considering the second necessary condition, we investigate the Fermi-level requirement for inducing large Berry curvature effects by SNLs or FNSs. Assume that a system hosts an SNL along the $(\pi,\pi,w)$ HSL and an FNS at the $(u,v,\pi)$ HSPL, as depicted by the red line and blue plane in Fig. \ref{strategy}(d), respectively. For simplicity, we take the SNL as an example to illustrate the requirement. The general bands structures of the SNL are plotted in Fig. \ref{strategy}(e). Due to the band dispersion along the SNL, these degenerate bands create a broad energy window. We select three representative k-points along the SNL (labeled $\bm k_1$, $\bm k_2$, $\bm k_3$) to detail their Berry curvature contributions, with the energies of $\bm k_1$, $\bm k_2$, $\bm k_3$ being below, exactly at, and above the Fermi level, respectively. At each k-point, we plot its band structures in the $k_x$-$k_y$ plane, which exhibits a Dirac cone. Near the $\bm k_1$ point, where the bands are almost fully occupied, which causes significant cancellation of Berry curvature between the upper and lower bands, resulting in relatively small Berry curvature in this region. Near the $\bm k_2$ point located at the Fermi surface, the bands are almost half-occupied, thus the Berry curvature of the upper and lower bands does not cancel, resulting in significant enhenced Berry curvature in this region. Near the $\bm k_3$ point, where the bands are nearly unoccupied, resulting in relatively small Berry curvature in this region. Therefore, along the SNL from $\bm k_1$ to $\bm k_3$ in the BZ, the Berry curvature magnitude $|\bm \Omega(\bm k)|$ first increases and then decreases, as illustrated by the black curve in Fig. \ref{strategy}(e). This behavior indicates that enhenced Berry curvature occurs near the intersection of the SNL and the Fermi surface, specifically at the $\bm k_2$ point. Even when a perturbation shifts the Fermi level, the SNL can still contribute significant Berry curvature as long as the Fermi level remains within its broad energy window. In this case, the peak of the black curve representing $|\bm \Omega(\bm k)|$ simply shifts along the SNL. Similarly, FNSs can intersect the Fermi surface forming lines within a broad energy window, thereby contributing significant Berry curvature. Thus, SNLs and FNSs can serve as robust sources of Berry curvature. Furthermore, for any necessarily existing SNL or FNS, all bands on it are degenerate, creating a series of such energy windows for large Berry curvature effects such as AHE, as shown in Fig. \ref{strategy}(f). By contrast, we also demonstrate accidental nodal points (NPs), which create a single, relatively narrow energy window for large AHC. Therefore, for any real material crystallizing in one of the 158 MSGs (see Table \ref{AHC_MSG}), its Fermi level has a high probability of lying within the energy window series, thereby giving rise to a large AHC.    \par

\begin{figure*}[htp]
	\centering
	{\includegraphics[width=1.0\textwidth]{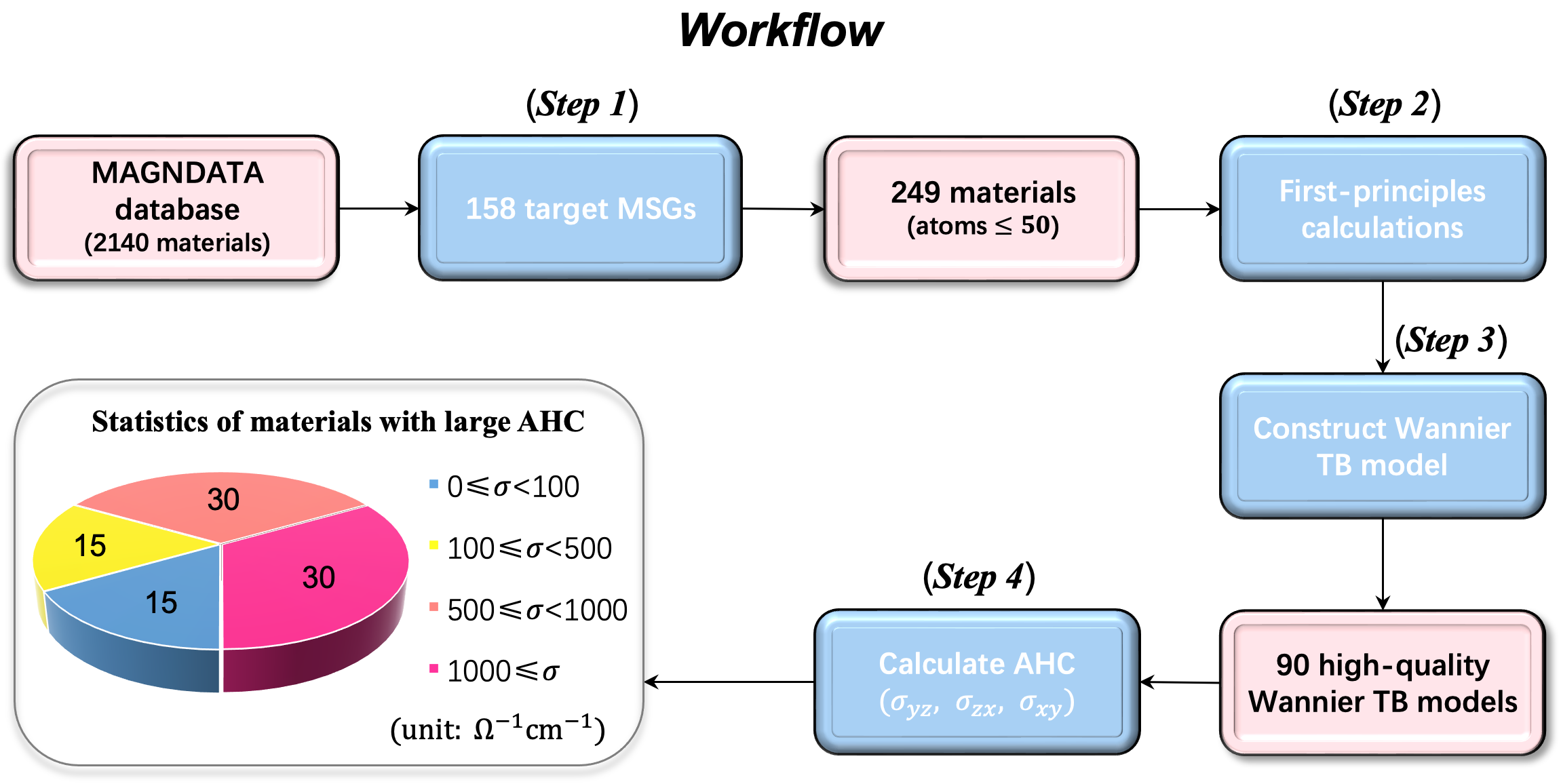}}
	\captionsetup{justification=raggedright}
	\caption{Workflow for identifying materials with large AHC via theory-driven high-throughput first-principles calculations. The workflow consists of four main steps: (1) Based on the 158 target MSGs listed in Table \ref{AHC_MSG} hosting FNSs, SNLs, or both and allowing for nonvanishing AHC, 277 magnetic materials are screened from the MAGNDATA magnetic material database. Among these, 249 materials with fewer than 50 atoms per primitive cell are selected for calculations. (2) High-throughput first-principles calculations are performed on the 249 materials. (3) An in-house developed code is employed to automatically construct 90 high-quality Wannier tight-binding (TB) models based on the first-principles results. (4) The AHC is calculated from these 90 Wannier TB models, followed by statistical analysis of the AHC magnitudes.  }
	\label{flow}
\end{figure*}

\begin{figure*}[htp]
	{\includegraphics[width=0.6\textwidth]{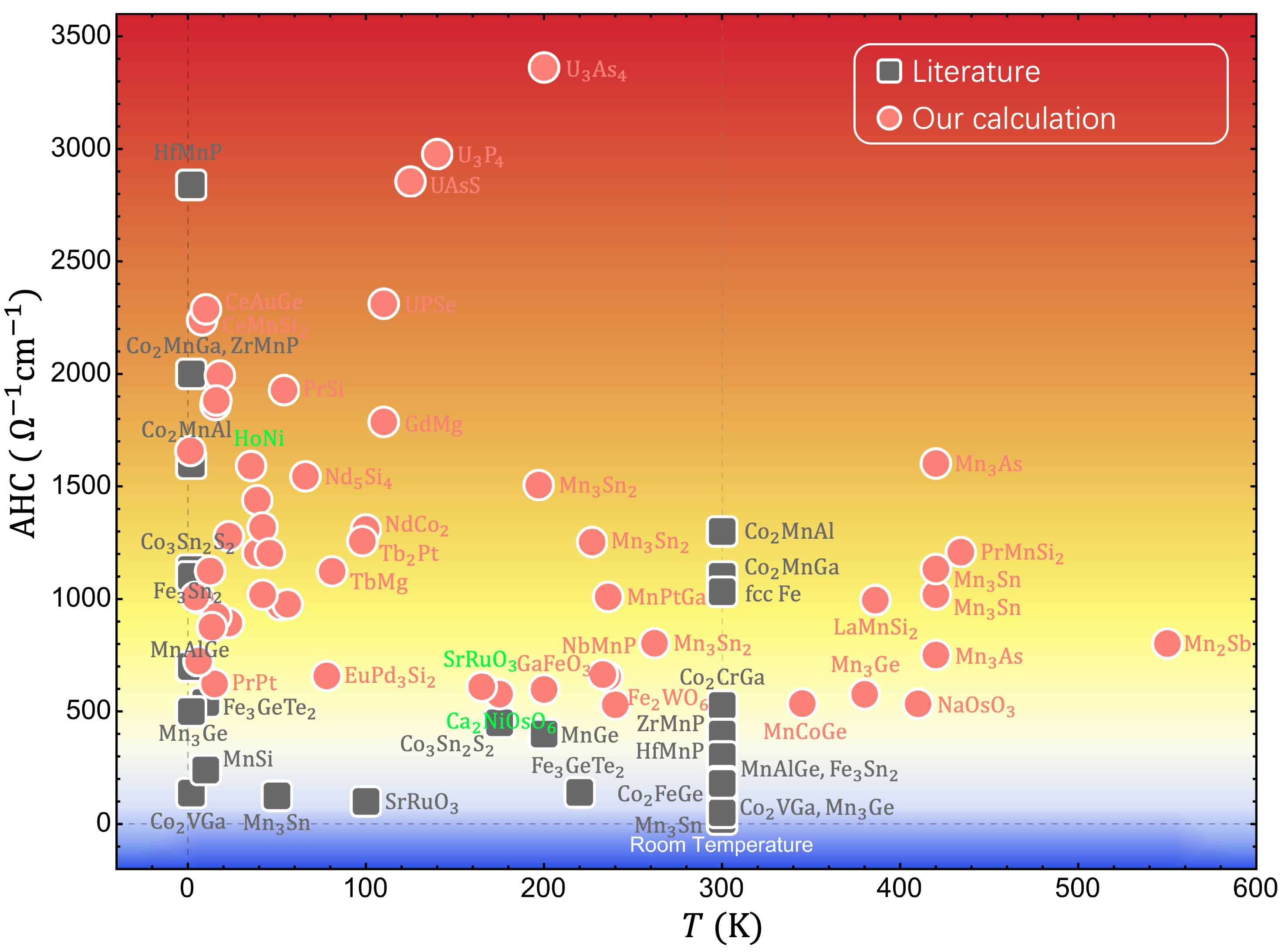}}
	\captionsetup{justification=raggedright}
	\caption{Comparison of AHC values and temperatures between our calculations and literature data. The 60 materials exhibiting AHC values exceeding $500\,\Omega^{-1}{\rm cm}^{-1}$ in this work are shown. The temperatures in our calculations and in the literature correspond to the transition and measurement temperatures, respectively. Materials highlighted in green are discussed in the main text.  }
	\label{matsfig}
\end{figure*}

\section{Theory-driven high-throughput calculations for magnetic materials with large AHC} \label{HTC}
Guided by our theoretical design strategy, we employ high-throughput calculations to identify materials with large AHC. The workflow consists of four main steps, as shown in Fig. \ref{flow}. 

\textit{Step 1}: Based on the 158 target MSGs that host FNSs, SNLs, or both and allow for nonvanishing AHC, as listed in Table \ref{AHC_MSG}, we screen out 277 matrials from Bilbao’s MAGNDATA magnetic material database~\cite{bilbao}, which contains 2140 experimentally synthesized magnetic materials. Before performing the calculations, the material structures are standardized: the working setting convention of each material is converted to a standard convention using the FINDSYM program~\cite{findsym-web, findsym-paper}, and the conventional cell is reduced to its primitive cell. Among these, 249 materials with fewer than 50 atoms in their primitive cell are selected for calculations.

\textit{Step 2}: First-principles calculations are performed for the 249 materials. For materials containing $d$- or $f$-electron elements, the DFT + $U$ approach is adopted. The Hubbard $U$ values for $d$- and $f$-electron elements are taken from Ref.~\cite{Ud-prm} and Ref.~\cite{Uf-prb}, respectively. 184 materials are selected, as their calculated magnetic moments are in good agreement with their experimental values, ensuring that the calculated ground states of these materials possess the correct MSG. Next, by calculating the density of states (DOS), 130 materials with band gaps smaller than $0.5\,\rm eV$ are selected for further calculations.

\textit{Step 3}: For these 130 materials, Wannier tight-binding (TB) models are automatically constructed based on the results of first-principles calculations. This step involves three main parts: (i) Projected electronic band structure calculations are performed to determine the orbital character of the bands; (ii) An in-house developed code is used to automatically select the suitable energy windows and projectors for generating the input files for WANNIER90 software~\cite{wannier90-jpc,wannier90-1-cpc,wannier90-2-cpc,wannier90-jpc}; (iii) Wannier TB models are then constructed using the WANNIER90 software. As a result, 90 high-quality Wannier TB models are obtained, each ensuring that the spread of all Wannier orbitals is below 10 $\rm \AA^2$. 

\textit{Step 4}: Based on the 90 high-quality Wannier TB models, the Fermi-level-dependent AHC is calculated within an energy range of $\pm2\,\rm eV$ around the Fermi level using the WANNIERTOOLS software~\cite{wanniertools-cpc}. Consequently, 75 materials exhibit AHC values exceeding $100\,\Omega^{-1}{\rm cm}^{-1}$, including 60 exceeding $500\,\Omega^{-1}{\rm cm}^{-1}$ and 30 exceeding $1000\,\Omega^{-1}{\rm cm}^{-1}$, as detailed in SM II~\cite{SM2}. In particular, the 60 materials with AHCs value exceeding $500\,\Omega^{-1}{\rm cm}^{-1}$ are compared with those previously reported in the literature, as shown in Fig. \ref{matsfig}. It is worth noting that, to the best of our knowledge, 59 of these 75 magnetic materials are newly proposed as candidates for exhibiting large AHC and have not been discussed previously.   \par 

Next, we select two magnetic materials, 0.732 $\rm SrRuO_3$ and 0.796 $\rm Ca_2NiOsO_6$ (newly proposed), as candidates to showcase their large AHC induced by the FNSs and SNLs, with one and two nonvanishing AHC components for each material, respectively. Note that the numbers preceding the chemical formulas correspond to their entries in the MAGNDATA database. \par

\begin{figure*}[ht]
	{\includegraphics[width=\textwidth]{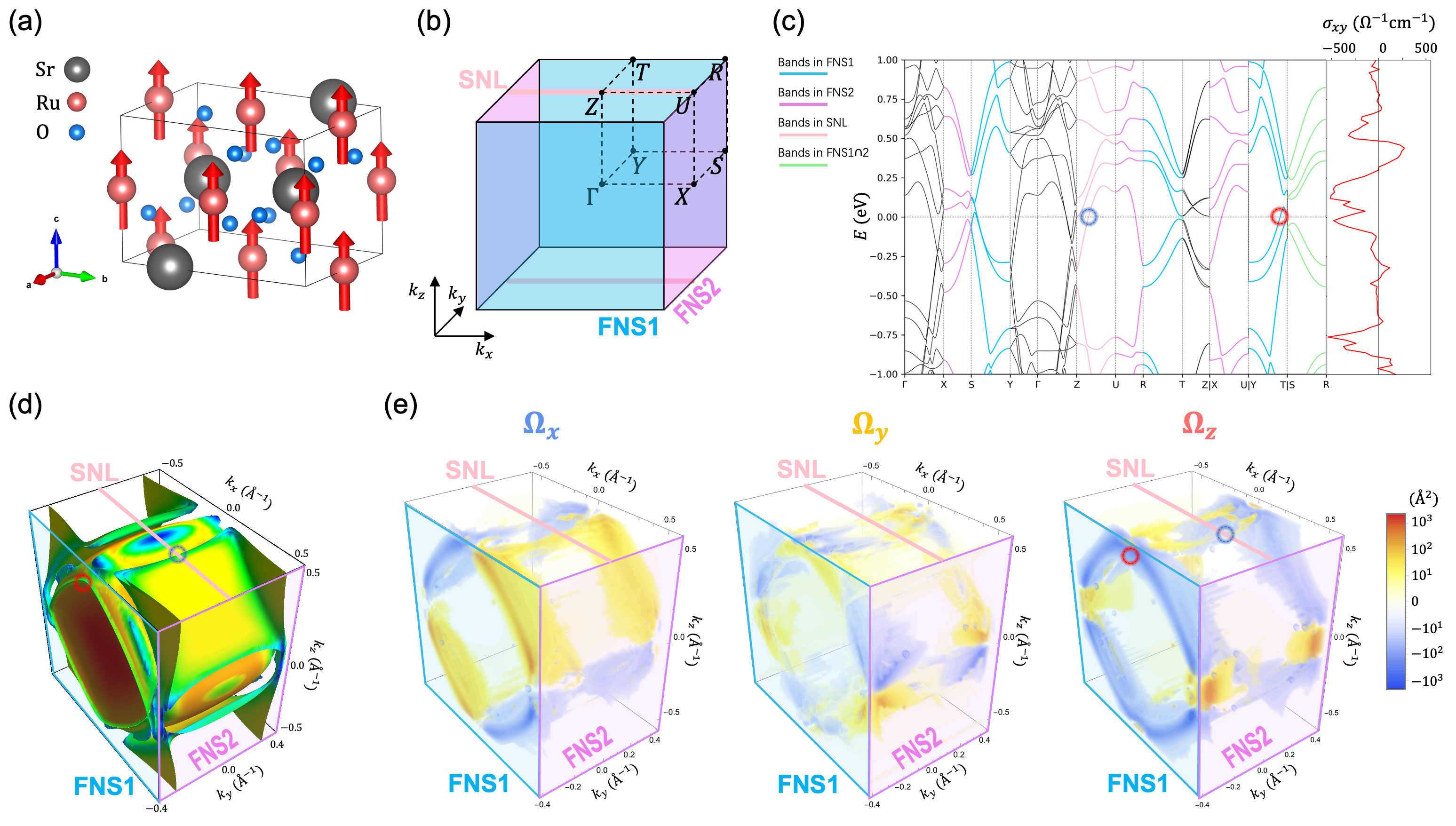}}
	\captionsetup{justification=raggedright}
	\caption{Calculation results for the magnetic material 0.732 $\rm SrRuO_3$. (a) Crystal and magnetic structures of $\rm SrRuO_3$, which crystallizes in type-III MSG 62.446. The magnetic moments of Ru atoms are represented by red arrows along the $c$ axis. (b) The extended nodal structures of MSG 62.446. The SNL, FNS1, and FNS2 are depicted as a pink line, a purple plane, and a blue plane in the BZ, respectively, coinciding with the $(0, w, \frac{1}{2})$ HSL, $(\frac{1}{2}, u, v)$ HSPL, and $(u, \frac{1}{2}, v)$ HSPL, where $\omega$, $u$, and $v$ are real parameters. (c) First-principles band structures and the Fermi-level-dependent AHC. The Fermi level is set to $0\,\rm eV$. The highlighted pink, purple, blue, and green bands correspond to the SNL, FNS1, FNS2, and the intersection of FNS1 and FNS2, respectively. The $\sigma_{xy}$ component, ranging from $-1\,\rm eV$ to $1\,\rm eV$, is shown alongside the band structures. (d) The Fermi surfaces intersecting FNS1, FNS2 and the SNL. (e) Distributions of the three Berry curvature components, $\Omega_x$, $\Omega_y$, and $\Omega_z$, in the BZ. Berry curvature hotspots are highlighted by red and blue dashed circles, which correspond to the red and blue dashed circles in (c) and (d), respectively.  }
	\label{SrRuO3}
\end{figure*}

\subsection{Candidate material $\rm SrRuO_3$ exhibiting large AHC} \label{candidate1}
The magnetic material 0.732 $\rm SrRuO_3$ is selected as a candidate to demonstrate the contributions of FNSs and SNLs to the AHC, due to its relatively simple band structure and Fermi surfaces, which facilitate the investigation of the origin of the AHC. $\rm SrRuO_3$ crystallizes in an orthorhombic lattice with type-III MSG 62.446. The crystal structure is shown in Fig. \ref{SrRuO3}(a). This material is a ferromagnet, with the magnetic moments of all $\rm Ru$ atoms aligned along the $c$-axis, thereby breaking TRS. According to the tabulation for type-III MSG 62.446 with double-valued representations in SM I~\cite{SM1}, the HSPLs at $(\frac{1}{2}, u, v)$ and $(u, \frac{1}{2}, v)$ host FNSs, while the HSL at $(0, \omega, \frac{1}{2})$ hosts an SNL, where $\omega$, $u$, and $v$ are real parameters, as shown in Fig. \ref{SrRuO3}(b). The coordinates of the SNLs and FNSs are given in fractional form based on the conventional basis vectors $(\vec{b}_1,\vec{b}_2,\vec{b}_3)$, as detailed in SM I~\cite{SM1}. The FNSs are represented by the purple and blue planes at the boundary of the BZ, labeled FNS1 and FNS2, respectively, while the SNLs are represented by the pink lines at the top and bottom. The first-principles band structure is shown in Fig. \ref{SrRuO3}(c), with the corresponding high-symmetry paths indicated by dashed lines in Fig. \ref{SrRuO3}(b). In the band structure, the highlighted pink, purple, blue, and green bands are two-fold degenerate and correspond to the SNL, FNS1, FNS2, and the intersection of FNS1 and FNS2, respectively. From Fig. \ref{SrRuO3}(c), it is evident that the bands associated with FNS1, FNS2 and the SNL cross the Fermi level. According to Table \ref{AHC_MSG}, the symmetry-adapted AHC tensor for MSG 62.446 is $\bm{\sigma}=(0, 0, \sigma_{xy})$, with only one nonvanishing AHC component, $\sigma_{xy}$. We then calculate the Fermi-level-dependent AHC based on the Wannier TB model, and the resulting $\sigma_{xy}$ curve is shown alongside the band structure in Fig. \ref{SrRuO3}(c). At the Fermi level, $\sigma_{xy}$ reaches approximately $500\,\Omega^{-1}{\rm cm}^{-1}$ and remains large over a broad energy window. Additionally, the other two components, $\sigma_{yz}$ and $\sigma_{zx}$, are vanishing (see SM II~\cite{SM2}), consistent with the symmetry constraints of MSG 62.446.  \par

Next, we investigate in detail the origin of the large AHC in $\rm SrRuO_3$. The calculated Fermi surfaces are shown in Fig. \ref{SrRuO3}(d), which appear relatively simple. For clarity, the SNLs and FNSs are also depicted in Fig. \ref{SrRuO3}(d) to illustrate their positions relative to the Fermi surfaces. It is evident that the Fermi surfaces intersect both FNS1 and FNS2, forming continuous curves widely distributed across the BZ, and intersect with the SNL at two points. The AHC components $\sigma_{yz}$, $\sigma_{zx}$, and $\sigma_{xy}$ correspond to the integrals of the Berry curvature components $\Omega_x$, $\Omega_y$, and $\Omega_z$, respectively, as expressed in Eq. (\ref{bck},\ref{ahc}). To identify the main contributors to the large AHC, we calculate the distributions of these three Berry curvature components, $\Omega_x$, $\Omega_y$, and $\Omega_z$, across the entire BZ, as shown in Fig. \ref{SrRuO3}(e). The calculated distributions manifest that regions with large $\Omega_x$, $\Omega_y$, and $\Omega_z$ are concentrated around the intersections of the Fermi surfaces with FNS1, FNS2, and the SNL. The Berry curvature magnitude near these intersections can reach up to $10^3\,\rm \AA^2$. To determine the origin of the large Berry curvature, we focus on a hotspot of $\Omega_z$ on FNS1, highlighted by a red dashed circle in the right panel of Fig. \ref{SrRuO3}(e). This hotspot lies at the intersection of the Fermi surfaces and FNS1, as indicated by the same marker in Fig. \ref{SrRuO3}(d). In the band structure, this hotspot corresponds to a k-point where the energy bands cross the Fermi level along the Y-T path, as also indicated by the same marker in Fig. \ref{SrRuO3}(c). Thus, this Berry curvature hotspot originates from FNS1. Similarly, the hotspot of $\Omega_z$ near the SNL, highlighted by a blue dashed circle in the right panel of Fig. \ref{SrRuO3}(e), originates from the SNL. Although the intersections of the Fermi surfaces and the SNL are points, the Fermi surfaces emanating from these points are nearly degenerate, resulting in lines with a small gap and substantial Berry curvature near them. Furthermore, the hotspots of $\Omega_z$ near FNS1 exhibit the same sign as the $\sigma_{xy}$ component of the AHC, indicating that the $\Omega_z$ originating from FNS1 positively contributes to $\sigma_{xy}$. In contrast, the significant $\Omega_z$ near the intersections of the Fermi surfaces with FNS2 and the SNL shows the opposite sign to $\sigma_{xy}$, indicating that the $\Omega_z$ originating from FNS2 and the SNL negatively contributes to $\sigma_{xy}$. Moreover, there are no other sources of Berry curvature apart from the FNSs and SNL. Therefore, we conclude that FNS1 is the primary contributor to the large AHC component $\sigma_{xy}$ at the Fermi level. Note that as the Fermi level shifts, the contributions of FNS1, FNS2, and the SNL may vary, potentially altering the primary contributor. Finally, although FNS1, FNS2, and the SNL also generate significant Berry curvature components $\Omega_x$ and $\Omega_y$, the corresponding AHC components $\sigma_{yz}$ and $\sigma_{zx}$ vanish due to symmetry constraints imposed by the MSG. For instance, the mirror symmetry $m_z$ ensures that $\Omega_x$ and $\Omega_y$ have opposite signs at two $m_z$ related k-points $(k_x, k_y, k_z)$ and $(k_x, k_y, -k_z)$, while $\Omega_z$ remains invariant. As a result, $\Omega_x$ and $\Omega_y$ are odd functions of $k_z$, and their integrals over the entire BZ cancel out, leading to the vanishing of $\sigma_{yz}$ and $\sigma_{zx}$. Interestingly, although the net $\Omega_x$ and $\Omega_y$ vanish, they form finite quadrupoles in the BZ, which can contribute a third-order NLAHE.   \par

\begin{figure*}[ht]
	{\includegraphics[width=\textwidth]{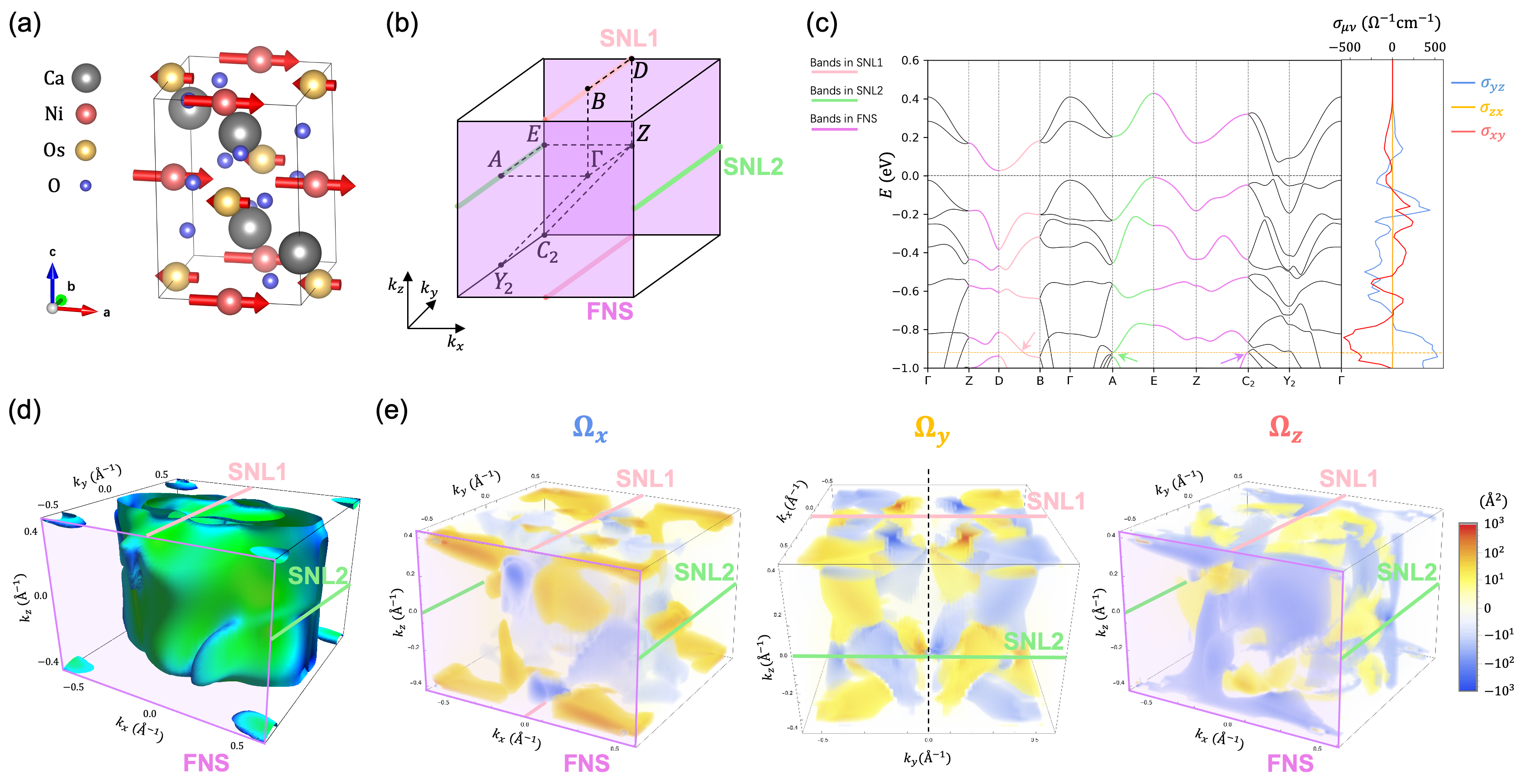}}
	\captionsetup{justification=raggedright}
	\caption{Calculation results for the magnetic material 0.796 $\rm Ca_2NiOsO_6$. (a) Crystal and magnetic structures of $\rm Ca_2NiOsO_6$, which crystallizes in type-III MSG 14.79. The magnetic moments are represented by red arrows along the $a$ axis. (b) The extended nodal structures of MSG 14.79. SNL1, SNL2, and the FNS are depicted as a pink line, a green line, and a purple plane in the BZ, respectively, coinciding with the $(0, \omega, \frac{1}{2})$ HSL, $(\frac{1}{2}, \omega, -\frac{1}{2})$ HSL, and $(u, \frac{1}{2}, v)$ HSPL, where $\omega$, $u$, and $v$ are real parameters. (c) First-principles band structures and the Fermi-level-dependent AHC. The Fermi level is set to $0\,\rm eV$. The highlighted pink, green, and purple bands correspond to SNL1, SNL2, and the FNS, respectively. The red (blue) curve of the $\sigma_{xy}$ ($\sigma_{yz}$) component, ranging from $-1\,\rm eV$ to $1\,\rm eV$, is shown alongside the band structures. The orange dashed line represents the set Fermi level of $E_{\rm F}=-0.92\, \rm eV$, with the bands crossing this set Fermi level on SNL1, SNL2, and the FNS indicated by pink, green, and purple arrows respectively. (d) The Fermi surfaces for $E_{\rm F}=-0.92\,\rm eV$ intersecting SNL1, SNL2, and the FNS. (e) Distribution of the three Berry curvature components, $\Omega_x$, $\Omega_y$, and $\Omega_z$, in the BZ for $E_{\rm F}=-0.92\,\rm eV$. }
	\label{Ca2NiOsO6}
\end{figure*}

\subsection{New candidate material $\rm Ca_2NiOsO_6$ exhibiting large AHC} \label{candidate2}
The magnetic material 0.796 $\rm Ca_2NiOsO_6$ is selected as a representative of the 59 new candidate materials to demonstrate the contributions of SNLs and FNSs to the AHC, due to its relatively simple band structure and Fermi surfaces, as well as the presence of two nonvanishing AHC components. The crystall structure of $\rm Ca_2NiOsO_6$ is shown in Fig. \ref{Ca2NiOsO6}(a). This material is a ferrimagnet, with the magnetic moments of all $\rm Ni$ and $\rm Os$ atoms aligned along the $a$-axis, thereby breaking TRS. This material crystallizes in a monoclinic lattice with type-III MSG 14.79. Although the lattice is monoclinic, its $\beta$ angle is $\rm 90.29^\circ$, which is very close to $\rm 90^\circ$, making the structure nearly orthorhombic. According to the tabulation of type-III MSG 14.79 with double-valued representations in SM I~\cite{SM1}, the HSLs at $(-\frac{1}{2}, 0, \omega)$ and $(\frac{1}{2}, \frac{1}{2}, \omega)$ host SNLs, while the HSPL at $(u, v, \frac{1}{2})$ hosts an FNS. For this mateial, there exists a unitary transformation between the basis vectors in our theory and those in the material, which cyclically permutes the basis vectors in our theory as $(\vec{a}_1, \vec{a}_2, \vec{a}_3)\rightarrow(\vec{a}_2, \vec{a}_3, \vec{a}_1)$. As a result, in the basis vectors of the material, the HSLs at $(0, \omega, \frac{1}{2})$ and $(\frac{1}{2}, \omega, -\frac{1}{2})$ host SNLs, and the HSPL at $(v, \frac{1}{2}, u)$ hosts an FNS, as shown in Fig. \ref{Ca2NiOsO6}(b). The SNLs are represented by the pink and green lines at the top (bottom) and right (left) of the BZ, labeled SNL1 and SNL2, respectively. The FNS is represented by purple plane in front of or behind the BZ, labeled FNS. The first-principles band structure is shown in Fig. \ref{Ca2NiOsO6}(c), and the corresponding high-symmetry paths are indicated by dashed lines in Fig. \ref{Ca2NiOsO6}(b). In the band structure, the highlighted pink, green, and purple bands are two-fold degenerate, and correspond to SNL1, SNL2, and the FNS, respectively. According to Table \ref{AHC_MSG}, the symmetry-adapted AHC tensor for MSG 14.79 is ${\bm \sigma}=(\sigma_{yz}, \sigma_{zx}, 0)$. Considering the transformation of the basis vectors, the $\sigma_{yz}$ and $\sigma_{xy}$ components are theoretically nonvanishing, namely ${\bm \sigma}=(\sigma_{yz}, 0, \sigma_{xy})$. Notably, although the basis vectors of a material may differ from our convention by a unitary transformation, which results in corresponding unitary transformation in the AHC tensor relative to our theoretical result, the number of the nonvanishing AHC components remains invariant. We then calculate the Fermi-level-dependent AHC of this material based on the Wannier TB model. The resulting AHC components curves are shown alongside the band structure in Fig. \ref{Ca2NiOsO6}(c). It is evident that the $\sigma_{yz}$ and $\sigma_{xy}$ components remain large over a broad energy window, both reaching $500\,\Omega^{-1}{\rm cm}^{-1}$ near $E_{\rm F}=-0.9\,{\rm eV}$, while the $\sigma_{zx}$ component is vanishing, consistent with the theoretical predictions for MSG 14.79. \par 

Similarly, we investigate in detail the origin of the large AHC in $\rm Ca_2NiOsO_6$. We set the Fermi level at $-0.92\,\rm eV$ (indicated by the orange dashed line in Fig. \ref{Ca2NiOsO6}(c)), where both $\sigma_{yz}$ and $\sigma_{xy}$ are relatively large, approximately 500 $\rm \Omega^{-1}{cm}^{-1}$ and $-400\,\Omega^{-1}{\rm cm}^{-1}$, respectively. We then calculate the Fermi surfaces at $E_{\rm F}=-0.92\,\rm eV$, as shown in Fig. \ref{Ca2NiOsO6}(d), which appear relatively simple. From both the Fermi surfaces and band structure, it is evident that SNL1, SNL2, and the FNS all intersect the Fermi surfaces. To identify the main contributors to the large AHC, we calculate the distributions of the three Berry curvature components, $\Omega_x$, $\Omega_y$, and $\Omega_z$, across the entire BZ, as shown in Fig. \ref{Ca2NiOsO6}(e). The calculated distributions show that large $\Omega_x$, $\Omega_y$, or $\Omega_z$ are distributed near the intersections of the Fermi surfaces with the FNS, SNL1, and SNL2. Moreover, the $\Omega_x$ and $\Omega_z$ originating from the FNS, SNL1 and SNL2 exhibit the same sign as the corresponding AHC components $\sigma_{yz}$ and $\sigma_{xy}$, respectively. Therefore, the FNS, SNL1 and SNL2 all positively contribute to the AHC components $\sigma_{yz}$ and $\sigma_{xy}$, making them the main contributors. Finally, although the FNS, SNL1 and SNL2 also generate significant $\Omega_y$, the AHC component $\sigma_{zx}$ vanishes due to symmetry constraints imposed by the MSG. Specifically, due to the $\Theta C_{2y}$ symmetry, the distribution of $\Omega_y$ is antisymmetric with respect to the $k_y=0$ plane, leading to the vanishing integral of $\Omega_y$ over the entire BZ, and thus, the vanishing of $\sigma_{zx}$. Interestingly, the $\Omega_y$ originating from SNL1, SNL2, and the FNS form quadrupoles in the BZ, which can contribute to a third-order NLAHE.  \par

\begin{figure*}[ht]
	{\includegraphics[width=\textwidth]{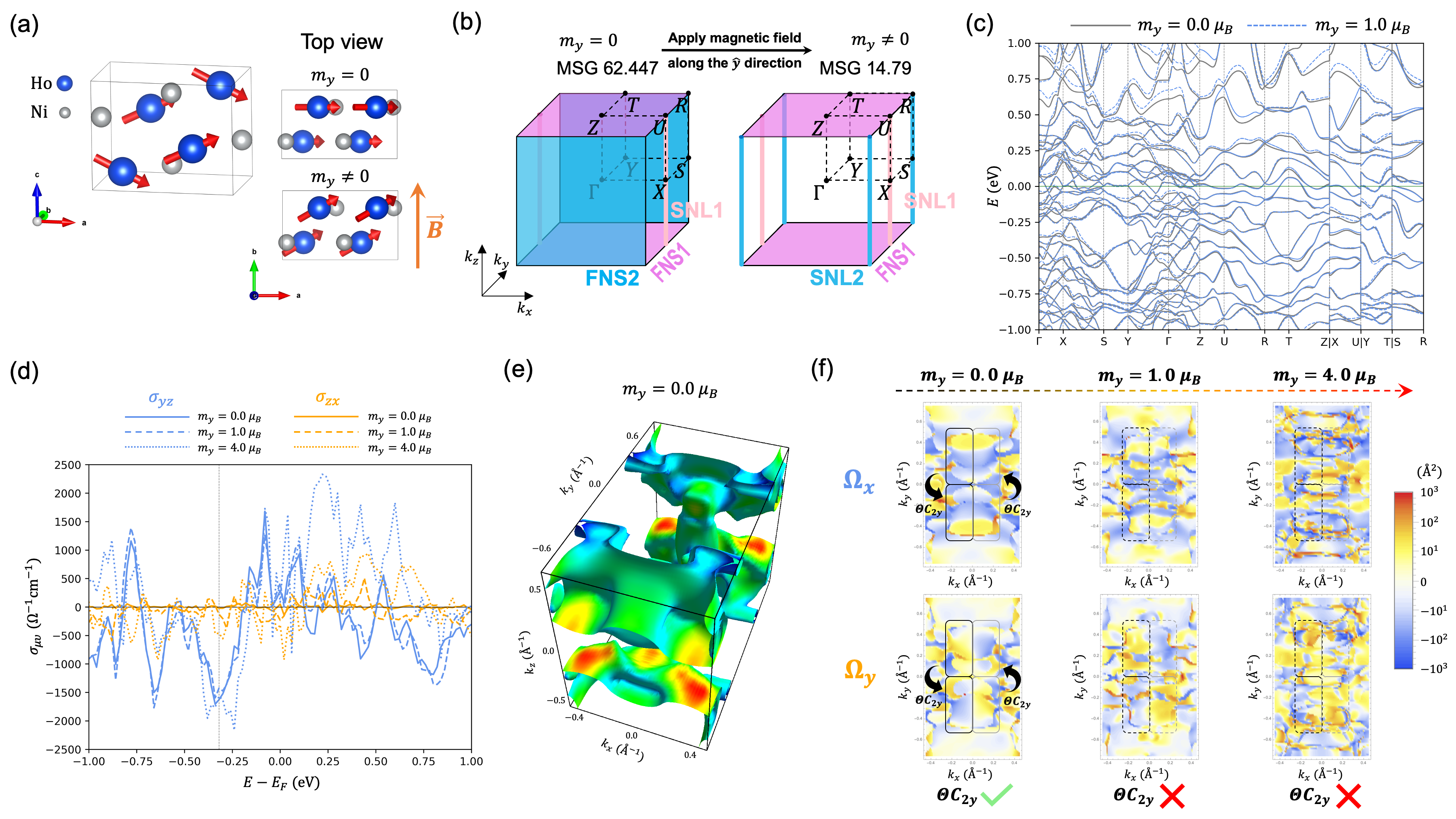}}
	\captionsetup{justification=raggedright}
	\caption{Tuning the AHC of the magnetic material 0.480 $\rm HoNi$ by an external magnetic field. (a) Crystal and magnetic structures of 0.480 $\rm HoNi$. The magnetic moments of $\rm Ho$ atoms are represented as red arrows lying in the $a$-$c$ plane. When an external magnetic field is applied along the $\hat{y}$ direction, the $m_y$ component of the magnetic moment becomes nonzero. (b) Evolution of nodal structures induced by the applied magnetic field. The MSG of the material changes from type-III 62.447 (left) to type-III 14.79 (right) upon application of the external magnetic field along the $\hat{y}$ direction, with the corresponding nodal structures shown in the left and right panels, respectively. The FNS1, FNS2, SNL1, and SNL2 are depicted as a purple plane, a blue plane, a pink line, and a blue line in the BZ, respectively, coinciding with $(u, v, \frac{1}{2})$ HSPL, $(u, \frac{1}{2}, v)$ HSPL, $(\frac{1}{2}, 0, \omega)$ HSL, and $(\frac{1}{2}, \frac{1}{2}, \omega)$ HSL, where $\omega$, $u$, and $v$ are real parameters. (c) First-principles band structures. The Fermi level is set to $0\,\rm eV$. The grey solid and blue dashed bands correspond to the bands for $m_y=0.0\,\mu_B$ and $m_y=1.0\,\mu_B$, respectively. (d) The Fermi-level-denpendent AHC components $\sigma_{yz}$ and $\sigma_{zx}$, ranging from $-1\,\rm eV$ to $1\,\rm eV$, are shown as the blue and orange curves, respectively. Solid, dashed, and dotted lines correspond to $m_y=0.0\,\mu_B$, $m_y=1.0\,\mu_B$, and $m_y=4.0\,\mu_B$, respectively. (e) The Fermi surfaces for $m_y=0.0\,\mu_B$ and $E_{\rm F}=-0.32\,\rm eV$ intersecting FNS1 and FNS2. (f) Distributions of the Berry curvature components, $\Omega_x$ and $\Omega_y$, on the $k_z=0.495\,b_3$ plane near FNS1 ($k_z=0.5\,b_3$) with $E_{\rm F}=-0.32\,\rm eV$. In the left panel, the Berry curvature within the two black (grey) boxes is related by $C_{2y}\Theta$ symmetry, resulting in the antisymmetry of $\Omega_y$ about $k_y=0$. In the middle and right panels, as $m_y$ increases, the Berry curvature within the two black (grey) dashed boxes shows increasing deviations from the $C_{2y}\Theta$ symmetry.  }
	\label{HoNi}
\end{figure*}

\section{Tuning AHC through symmetry breaking} \label{tune}
Based on the comprehensive tabulation of FNSs and SNLs, along with the symmetry-adapted AHC tensors in the 1651 MSGs discussed in Sec. \ref{FNSSNL}, we identify 807 MSGs that host FNSs, SNLs, or both but exhibit vanishing AHC, with double-valued representations. While these FNSs or SNLs can also generate substantial Berry curvature, the AHC vanishes due to cancellation induced by certain symmetries in the MSGs. However, this scenario differs from a trivial zero AHC, which occurs when there is a lack of significant Berry curvature. In this scenario, if certain symmetries of the system are broken by methods such as applying an external field, the FNSs or SNLs can contribute finite Berry curvature, leading to nonvanishing AHC. In this regard, the MSG of the system will be reduced to a subgroup that allow for nonvanishing AHC. Therefore, the 807 MSGs provide a fertile platform for investigating the tuning of AHC through symmetry breaking. We then examine these 807 MSGs to identify those with subgroups that allow for nonvanishing AHC while still hosting FNSs, SNLs or both. In other words, we aim to identify the MSGs that have subgroups belonging to the 158 MSGs listed in Table \ref{AHC_MSG}. As a result, 583 MSGs are identified, and the complete evolution paths of these subgroups are provided in SM I~\cite{SM1}. Notably, among these 583 MSGs, 158 are type-III MSGs, which correspond to magnetic systems that can be manipulated by external magnetic fields to break certain symmetries, as detailed in SM I~\cite{SM1}. Furthermore, this approach is applicable not only for tuning the AHC from vanishing to nonvanishing through symmetry breaking, but also for tuning the AHC from having a single nonvanishing component to two. Consequently, 138 type-III MSGs are identified with one nonvanishing AHC component and these have subgroups with two nonvanishing AHC components, as also shown in SM I~\cite{SM1}. All 7 type-III MSGs with two nonvanishing AHC components in Table \ref{AHC_MSG} can be obtained through symmetry breaking from the 138 MSGs with one nonvanishing AHC component.    \par

Next, we take the magnetic material 0.480 $\rm HoNi$ as an example to demonstrate the tuning of the AHC by an external magnetic field. The crystal and magnetic structure of $\rm HoNi$ are shown in Fig. \ref{HoNi}(a). This material is a non-collinear ferromagnet, with the magnetic moments of all $\rm Ho$ atoms lying within the $x$-$z$ plane, and their component $m_y=0\,\mu_B$. This material crystallizes in type-III MSG 62.447, which hosts one SNL and two FNSs, as shown in the left panel of Fig. \ref{HoNi}(b). When an external magnetic field is applied along the $\hat{y}$ direction, the magnetic moments of the Ho atoms will exhibit a finite $m_y$ component. As a result, the MSG of the material transitions to type-III 14.79, which hosts two SNLs and one FNS, as shown in the right panel of Fig. \ref{HoNi}(b). FNS2 in MSG 62.447 is protected by the $\Theta C_{2y}$ symmetry, which is broken by the finite magnetic moment component $m_y$ in MSG 14.79. The first-principles band structures for $m_y=0.0\,\mu_B$ and $m_y=1.0\,\mu_B$ are compared in Fig. \ref{HoNi}(c). Along the S-Y and Y-T paths that lie on FNS2 in MSG 62.447, the bands for $m_y=0.0\,\mu_B$ are two-fold degenerate, whereas the bands for $m_y=1.0\,\mu_B$ are gapped, indicating that FNS2 is lifted when $m_y=1.0\,\mu_B$. Additionally, the bands along the S-R path at the boundary of FNS2 remain degenerate, leading to the formation of a new SNL, denoted as SNL2. Thus, the evolution of nodal structures in the calculated band structures is in good agreement with our theoretical predictions. According to Table \ref{AHC_MSG}, the symmetry-adapted AHC tensor for MSG 62.447 is $(\sigma_{yz}, 0, 0)$, while that for MSG 14.79 is $(\sigma_{yz}, \sigma_{zx}, 0)$. Therefore, when an external magnetic field is applied along the $\hat{y}$ direction, the $\sigma_{zx}$ becomes nonvanishing. We then calculate the Fermi-level-dependent AHC based on the Wannier TB model for $m_y=0.0\,\mu_B$, $m_y=1.0\,\mu_B$, and $m_y=4.0\,\mu_B$, as shown in Fig. \ref{HoNi}(d). The AHC component $\sigma_{yz}$ reaches $1500\,\Omega^{-1}{\rm cm}^{-1}$ in the $\pm1\,\rm eV$ range around the Fermi level for all three cases, and for $m_y=4.0\,\mu_B$, it exceeds $2000\,\Omega^{-1}{\rm cm}^{-1}$. The AHC component $\sigma_{zx}$ is zero for $m_y=0.0\,\mu_B$. As $m_y$ increases from $0.0\,\mu_B$ to $1.0\,\mu_B$ and $4.0\,\mu_B$, the maximum of $\sigma_{zx}$ increases from nearly $0\,\Omega^{-1}{\rm cm}^{-1}$ to over $100\,\Omega^{-1}{\rm cm}^{-1}$ and to over $500\,\Omega^{-1}{\rm cm}^{-1}$, respectively. To investigate the mechanism underlying the tuning of the AHC by the applied magnetic field, we set $E_{\rm F}=-0.32\,\rm eV$ where both $\sigma_{yz}$ and $\sigma_{zx}$ are relatively large, and then calculate the Fermi surfaces for $m_y=0.0\,\mu_B$, as shown in Fig. \ref{HoNi}(e). It is evident that both FNS1 and FNS2 intersect the Fermi surfaces for $m_y=0.0\,\mu_B$. We then plot the distributions of the Berry curvature components $\Omega_x$ and $\Omega_y$ on the $k_z=0.495\,b_3$ plane, which is close to FNS1 ($k_z=0.5\,b_3$), for $m_y=0.0\,\mu_B$, $m_y=1.0\,\mu_B$, and $m_y=4.0\,\mu_B$, as shown in Fig. \ref{HoNi}(f). At first glance, large Berry curvature components $\Omega_x$ and $\Omega_y$ appear near the intersetion of FNS1 and the Fermi surfaces, with magnitude reaching $\rm 10^3\,\AA^2$. For $m_y=0.0\,\mu_B$, as shown in the left panel of Fig. \ref{HoNi}(f), $\Omega_y$ is antisymmetric with respect to the $k_y=0$ plane due to the $\Theta C_{2y}$ symmetry in MSG 62.447. This antisymmetry causes the integral of $\Omega_y$ over the BZ to vanish, resulting in a vanishing $\sigma_{zx}$. In contrast, for $m_y=1.0\,\mu_B$, as shown in the middle panel of Fig. \ref{HoNi}(f), due to the breaking of the $\Theta C_{2y}$ symmetry, the distribution of $\Omega_y$ slightly deviates from the antisymmetry with respect to the $k_y=0$ plane. This deviation makes the integral of $\Omega_y$ over the BZ finite, thus inducing a nonvanishing $\sigma_{zx}$. Similarly, for $m_y=4.0\,\mu_B$, as shown in the right panel of Fig. \ref{HoNi}(f), the larger value of $m_y$ amplifies the deviation from antisymmetry, leading to a larger $\sigma_{zx}$. Note that for $m_y=4.0\,\mu_B$, the magnetic configuration of material 0.480 $\rm HoNi$ matches that of material 0.481 $\rm HoNi$ in the database at temperatures below 20 $\rm K$. This suggests that the same symmetry breaking can also be achieved by tuning the temperature.        \par

\section{Method} \label{method}
The first-principles calculations in this work are based on the density functional theory as implemented in the Vienna Ab-initio Simulation Package (VASP)~\cite{vasp1-kresse-prb, vasp2-kresse-cms}, using the projector augmented wave method~\cite{paw-kresse-prb}. The generalized gradient approximation~\cite{gga-perdew-prl}, in the Perdew-Burke-Enzerhof form, is adopted for the exchange and correlation functional~\cite{pbe-perdew-prl}. The Hubbard $U$ values for $d$- and $f$-electron elements are taken from Ref.~\cite{Ud-prm} and Ref.~\cite{Uf-prb}, respectively. The plane-wave energy cutoff is set to 1.5 times the default cutoff energy in all calculations. For self-consistent electronic ground-state calculations, BZ sampling is performed using the Gamma-centered Monkhorst-Pack grid, generated by the VASPKIT package~\cite{vaspkit-wang-cpc}, with a density of 0.02  $2\pi/\rm \AA$. The convergence criterion for electric self-consistent iterations is set to $10^{-7}\,\rm eV$. The Wannier TB model is constructed using the Maximally localized Wannier functions method~\cite{MLWF-1-prb,MLWF-2-prb}, as implemented in the WANNIER90 software~\cite{wannier90-jpc,wannier90-1-cpc,wannier90-2-cpc}. The Berry curvature at a single k-point is evaluated within the linear response theory of Kubo formalism, based on the Wannier TB model, as shown in Eq. (\ref{bcforband},\ref{bcfork}), where $n$ and $n^\prime$ denote the band indices, ${\rm Im}$ denotes taking the imaginary part, and $f(E_n(\bm{k})-E_{\rm F})$ is the Fermi-Dirac distribution function. The Berry curvature distribution over the entire BZ is then calculated using Eq. (\ref{bcforband},\ref{bcfork},\ref{bck}), with a $101\times101\times101$ k-point grid. The Fermi-level-dependent AHC is calculated using Eq. (\ref{bcforband},\ref{bcfork},\ref{ahc}) implemented in the WANNIERTOOLS software~\cite{wanniertools-cpc}, also with a $101\times101\times101$ k-point grid.  \par

\begin{equation}\label{bcforband}
	\Omega_{\mu \nu}^n(\bm k)=-2\,{\rm Im}\sum_{n^\prime\neq n}\frac{\langle u_n | \partial_\mu H|u_{n^\prime}\rangle \langle u_{n^\prime} | \partial_\nu H|u_n\rangle}{(E_n-E_{n^\prime})^2}
\end{equation}
\begin{equation}\label{bcfork}
	\Omega_{\mu \nu}(\bm k)=\sum_{n}\Omega_{\mu \nu}^n(\bm k)f(E_n({\bm k})-E_{\rm F})
\end{equation}

\begin{equation}\label{bck}
	\Omega_{c}(\bm k)=\varepsilon_{abc}\Omega_{ab}(\bm k)
\end{equation}

\begin{equation}\label{ahc}
	\sigma_{\mu \nu}=\frac{e^2}{\hbar VN_k}\sum_{\bm k}\Omega_{\mu \nu}({\bm k})
\end{equation}

\section{Conclusion and discussion} \label{conclusion}
In conclusion, we propose a rational design strategy for achieving large Berry curvature effects by extended SNLs and FNSs. We exhaustively tabulate all possible SNLs and FNSs in the 1651 MSGs. Based on this tabulation, we construct effective Hamiltonians for selected SNLs and FNSs, demonstrating their advantages in generating widely distributed Berry curvature and broad energy windows for large Berry curvature effects such as AHE. As an application, we investigate the symmetry-adapted AHC tensors for all 1651 MSGs and highlight 158 MSGs that host FNSs, SNLs, or both and allow for nonvanishing AHC. Guided by these 158 MSGs, we screen materials from the MAGNDATA magnetic material database for high-throughput first-principles calculations, and identify 75 materials with AHC values exceeding $100\,\Omega^{-1}{\rm cm}^{-1}$, including 60 exceeding $500\,\Omega^{-1}{\rm cm}^{-1}$ and 30 exceeding $1000\,\Omega^{-1}{\rm cm}^{-1}$.  It is worth noting that, to the best of our knowledge, 59 of these 75 magnetic materials are newly proposed as candidates for exhibiting large AHC. Specifically, we use the known candidate material $\rm SrRuO_3$ and the newly proposed candidate material $\rm Ca_2NiOsO_6$ to demonstrate the contributions of FNSs and SNLs to one and two nonvanishing AHC components, respectively. Notably, the Kagome ferromagnet $\rm Fe_3Ge$ was recently experimentally synthesized and identified to host an FNS at the $k_z=\pi$ HSPL, resulting in a large AHC value exceeding $1500\ (1300)\,\Omega^{-1}{\rm cm}^{-1}$ at $160\ (300)\,\rm K$~\cite{Fe3Ge-AHE}. Furthermore, we investigate the tuning of AHC through symmetry breaking, and identify 583 MSGs that host FNSs, SNLs, or both but exhibit vanishing AHC, which are applicable for AHC tuning through symmetry breaking. As an example, we show that the AHC of HoNi can be significantly tuned by applying an external magnetic field. We also identify Berry curvature quadrupoles in the candidate materials $\rm SrRuO_3$ and $\rm Ca_2NiOsO_6$, indicating that our strategy can be can be easily generalized to Berry curvature multipole effects, such as NLAHE. We hope that our strategy will guide both the theoretical and experimental design of materials with large Berry curvatures, thereby contributing to technological advancements in a wide range of device applications such as high-density, low-energy-consumption memory devices, and high-sensitivity sensors.    \par

\section{Acknowledgement}
This paper was supported by the National Natural Science Foundation of China (NSFC) under Grants No. 12188101, No. 12322404, No. 12104215, the National Key R\&D Program of China (Grant No. 2022YFA1403601), Innovation Program for Quantum Science and Technology, No. 2021ZD0301902, and Natural Science Foundation of Jiangsu Province (Grant No. BK20233001, BK20243011). F.T. was also supported by the Young Elite Scientists Sponsorship Program by the China Association for Science and Technology. X.W. also acknowledges support from the Tencent Foundation through the XPLORER PRIZE.

\section{References}

\bibliography{ref250122}

\end{document}